\documentclass[a4paper]{jpconf}
\usepackage{graphicx}

\def\dslash{/\kern-.1em \partial}

\def\al{\alpha}
\def\be{\beta}
\def\ga{\gamma}

\def\ep{\epsilon}

\def\la{\lambda}

\def\si{\sigma}

\def\La{\Lambda}

\def\mn{{\mu\nu}}

\def\frac#1#2{{\textstyle{{#1}\over {#2}}}}

\def\lsim{\mathrel{\rlap{\lower4pt\hbox{\hskip1pt$\sim$}}
    \raise1pt\hbox{$<$}}}
\def\gsim{\mathrel{\rlap{\lower4pt\hbox{\hskip1pt$\sim$}}
    \raise1pt\hbox{$>$}}}
\def\sqr#1#2{{\vcenter{\vbox{\hrule height.#2pt
         \hbox{\vrule width.#2pt height#1pt \kern#1pt
         \vrule width.#2pt}
         \hrule height.#2pt}}}}
\def\square{\mathchoice\sqr66\sqr66\sqr{2.1}3\sqr{1.5}3}

\newcommand{\beq}{\begin{displaymath}}
\newcommand{\eeq}{\end{displaymath}}
\newcommand{\bea}{\begin{eqnarray*}}
\newcommand{\eea}{\end{eqnarray*}}

\begin{document}
\title{Quantization of Space-like States in Lorentz-Violating Theories}

\author{Don Colladay}

\address{New College of Florida, Sarasota, FL 34234, USA}

\ead{colladay@ncf.edu}

\begin{abstract}
Lorentz violation frequently induces modified dispersion relations that can yield 
space-like states that impede the standard quantization procedures.  
In certain cases, an extended hamiltonian formalism can 
be used to define observer-covariant normalization factors for field expansions and phase space 
integrals.  These factors extend the theory to include non-concordant frames in which there 
are negative-energy states.  This formalism provides a rigorous way to quantize certain 
theories containing space-like states and allows for the consistent computation of Cherenkov
radiation rates in arbitrary frames and avoids singular expressions.
\end{abstract}

\section{Introduction}
\begin{figure}[h]
\begin{minipage}{18pc}
\includegraphics[width=18pc]{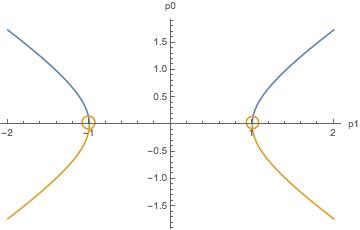}
\caption{\label{tachnu1}Plot of dispersion relation with separation between particle and anti-particle
states.}
\end{minipage}\hspace{2pc}%
\begin{minipage}{18pc}
\includegraphics[width=18pc]{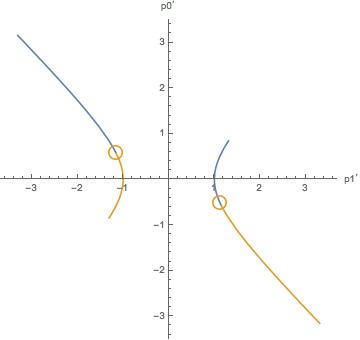}
\caption{\label{tachnu2}Plot of dispersion relation in boosted frame where some of 
particle states have $p_0^\prime < 0$.}
\end{minipage} 
\end{figure}

Theories involving space-like solutions to the dispersion relation yield negative-energy 
states that are usually so problematic that a consistent quantum field theory is elusive.
For example, a simple tachyonic neutrino model \cite{chodos} in which 
$$
L = i \overline \psi \ga_5 \not \partial \psi - m_\nu \overline \psi \psi,
$$
leads to a dispersion relation $p^2 = - m_\nu^2$, was proposed to model anomalous
neutrino mass measurement results.
Conventional attempts at quantization typically fail due to the impossibility of finding a 
reference frame-independent separation of particle and anti-particle states.
This leads to inconsistencies in the usual re-interpretation of the negative-energy states
in terms of anti-particles.
The situation is demonstrated in Figures \ref{tachnu1} and \ref{tachnu2} which shows the 
dispersion relation plotted in two different reference frames.  
Note that some of the anti-particle states in one frame are transformed into particle states in the 
other frame.

Similar issues can occur in Lorentz-violating theories, such as in the CPT-violating, massive
photon model (with $k_{AF}^\mu$ a constant background vector field)
\beq
\mathcal{L}_A = -\frac14 F_{\mu\nu} F^{\mu\nu} +
\frac12 k_{AF}^\kappa \epsilon_{\kappa\lambda\mu\nu}A^\lambda F^{\mu\nu}  
+ \frac12 m_\gamma^2 A_\mu A^\mu - {1 \over 2 \xi}(\partial_\mu A^\mu)^2,
\eeq
with momentum-space solution containing the perturbed dispersion relation factor 
\beq
R_T (p) ={1 \over 4} (p^2-m_\gamma^2)^2 - (p\cdot k_{AF})^2 + p^2 k_{AF}^2 = 0,
\eeq
which has the observer-covariant factorization $R_T(p) = R_+(p) R_- (p)$ with
\beq
R_\pm(p) = {1 \over 2} (p^2 - m_\gamma^2) \pm \sqrt{(p\cdot k_{AF})^2 - p^2 k_{AF}^2}.
\eeq

This factorization can be directly related to solutions to the modified
Dirac equation "off-shell" spinor solutions
\beq
{1 \over 2}(\not p - m - b_\mu \ga_5 \ga^\mu ) u_\pm(p) = {\cal R}_\pm(p) u_\pm( p),
\eeq
with ${\cal R}_\pm(p) = {1 \over 2} (p^2 - m^2 - b^2) \pm  \sqrt{(b \cdot p)^2 - b^2 p^2}$ as above
acting on spinors $u_\pm = (\not p + m - \ga_5 \not b) w_\pm$ giving the condition
\beq
{1 \over 2} \ep_{\mu\nu\alpha\be}\si^{\mn} p^\al b^\be w_\pm = \pm  \sqrt{(b \cdot p)^2 - b^2 p^2} w_\pm,
\eeq
which demonstrates that they are eigenstates of the Pauli-Lubanski vector.

The factor $R_+ (p) = 0$ has space-like solutions with $p^2 < 0$ at high-momenta.
\begin{figure}[h]
\begin{minipage}{18pc}
\includegraphics[width=18pc]{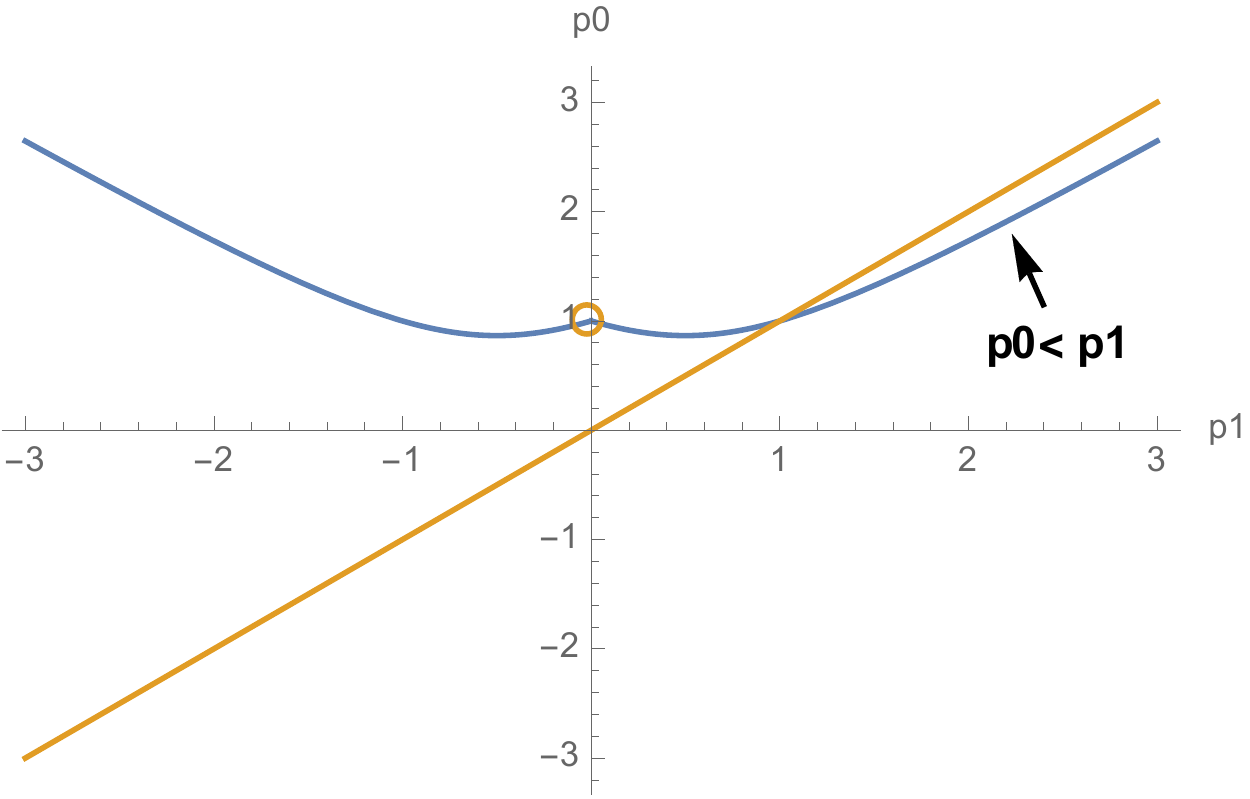}
\caption{\label{spacelike}Space-like states for dispersion
relation $R_+(p) = 0$ compared to conventional massless photon.}
\end{minipage}\hspace{2pc}%
\begin{minipage}{18pc}
\includegraphics[width=18pc]{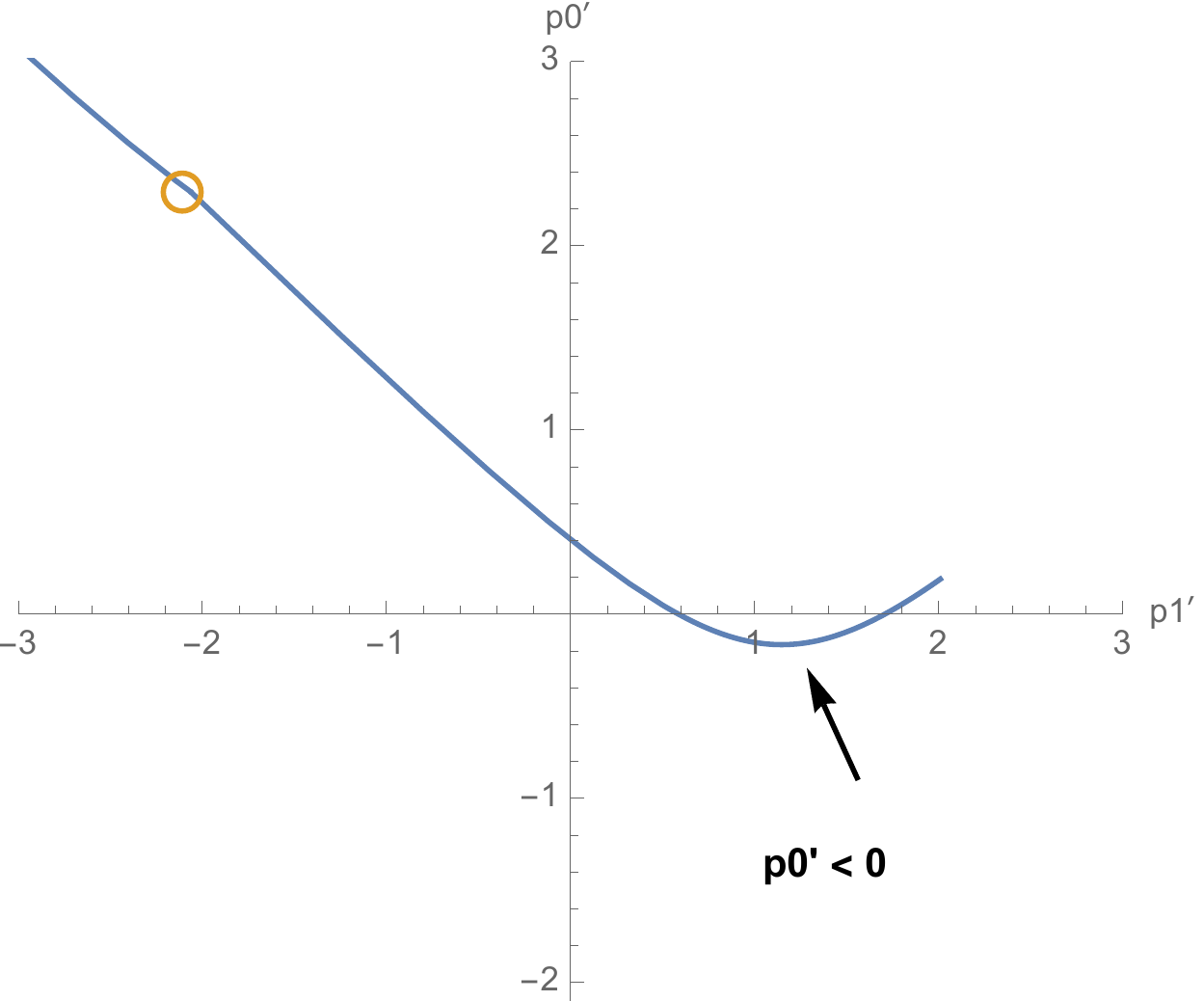}
\caption{\label{nege}Same states in a highly boosted frame where some energies are negative.}
\end{minipage} 
\end{figure}
Figure \ref{spacelike} shows the positive-energy particle states for the case of a pure
timelike $b^\mu$ coefficient.  Space-like states exist in the region where the dispersion
relation plot dips below the light-cone.  Figure \ref{nege} displays the energy and momentum
in a highly boosted frame where the energy dips below zero in some region.  
A first attempt may be to re-interpret these negative-energy solutions as anti-particles
and use conventional field normalization factors, but it 
turns out that this procedure leads to singular factors in the commutation relations that 
makes the quantization procedure suspect.  
Instead, an alternative procedure which makes use of an extended hamiltonian formalism will
be used that makes all of the states into particle states with an observer covariant interpretation 
of particle states that is valid in all frames.  
To identify the appropriate normalization factors, 
use is made of the classical mechanical limit of the theory.

\section{Classical Mechanics lagrangians in the SME}
The traditional method of computing lagrangians in the SME \cite{kosruss} uses a Legendre
transformation in conjunction with the group velocity to obtain the appropriate lagrange functions
from the dispersion relation.  
The process can be made to appear covariant by introducing an arbitrary path parametrization
and a four-velocity $u^\mu = d x^\mu / d \la$.
For example, the fermion dispersion relation with an external $b^\mu$-parameter considered
in \cite{kosruss} leads to a
dispersion relation isomorphic to the massive, CPT-violating photon one by making the replacements 
$b^\mu \rightarrow k_{AF}^\mu$, and $m^2 \rightarrow m_\ga^2 - k_{AF}^2$,
which preserves hermiticity provided $m_\ga^2 > k_{AF}^2$.
The relevant dispersion relation is
\beq
{\cal R}(p) = {1 \over 4} (p^2 - m^2 + b^2)^2 - (b \cdot p)^2 + m^2 b^2 = 0.
\eeq
The group velocity $\vec v = d \vec x / d t$ is computed using implicit differentiation as
\beq
{v_i} = {\partial p_0 \over  \partial p^i},
\eeq
which is well-defined away from any singular points where the energy surfaces become degenerate.
The relevant lagrangians are computed using the Legendre transformation $L = \vec  u \cdot \vec p - p^0(\vec p)$.
The expression for $\vec v (\vec p)$ is inverted for $\vec p(\vec v)$ to give
\beq
L_{\pm}[\vec v, x] = -m \sqrt{1 - \vec v^2} \mp \sqrt{(b^0 - \vec b \cdot \vec v)^2 - b^2 (1 - \vec v^2)},
\eeq
where the signs represent two valid solutions that reduce to the standard case when 
$b^\mu \rightarrow 0$.

The above procedure can be made explicitly covariant by introducing a zero-component 
of the four-velocity, $u^0(\la)$, and an 
arbitrary parametrization $\la(t)$ so that the action appears covariant in terms of the
four-velocity $u^\mu = d x^\mu  / d \la$, as
\beq
L_{\pm}[u^\mu, x] = -m \sqrt{u^2} \mp \sqrt{(b \cdot u)^2 - b^2 u^2}.
\eeq
The above approach has some issues which make a relativistic hamiltonian description difficult.
For one, the function $u^\mu$ has a "gauge" degree of freedom due to the
re-parametrization invariance and therefore it is not fully determined by the equation
of motion.
Computing the momentum $p^\mu = - {\partial L \over \partial u_\mu}$ gives
\beq
p^\mu = {m u^\mu \over \sqrt{u^2}} \pm 
{(u \cdot b) b^\mu - b^2 u^\mu \over \sqrt{(b \cdot u)^2 - b^2 u^2}},
\eeq
which is insensitive to a scaling of the four-velocity.  
This means that it is not invertible to find $u(p)$ unless some condition on $u^\mu$ is
enforced (such as $u^2 = 1$ when $\la = \tau$ is taken as the proper time).
Calculation of the relativistic hamiltonian yields
$$
{\cal H} = p^0 u^0 - \vec p \cdot \vec u - L = 0,
$$
(since $L = - u_\mu p^\mu$) yielding no useful hamiltonian formalism.
It is also unclear how $L_+$ and $L_-$ relate to solutions of $R_+ (p) = 0$ and $R_-(p) = 0$.

\section{Extended Hamiltonian Formalism} 
The relativistic hamiltonian can be extended to "off-shell" values using an extended hamiltonian
formalism originally due to Dirac.  
The procedure is to introduce a new variable $e(\la)$ as a lagrange multiplier \cite{colplb}
\beq
S^*_\pm = - \int \left[ m e^{-1} u^2 \pm \sqrt{(b\cdot u)^2 - b^2 u^2} -
{e \over m} {\cal R}_\mp (p,x) \right ] d \lambda, 
\eeq
with the observer covariant factorization
\beq
{\cal R}(p) = {\cal R}_+(p) {\cal R}_-(p),
\eeq
and
\beq
{\cal R}_\pm = {1 \over 2} \left( p^2 - m^2 - b^2 \right) \pm \sqrt{(b \cdot p)^2 - b^2 p^2},
\eeq
which agrees with the previous action "on-shell" where ${\cal R}_\pm = 0$, but modified action 
applies for unconstrained variations of $u^\mu$ and $p^\mu$ yielding an extended
relativistic hamiltonian 
\beq
{\cal H}^*_\pm =  -{e \over m}{\cal R}_\mp (p,x)
= - {e \over 2 m} \left( p^2 - m^2 - b^2 \pm 2 \sqrt{(b \cdot p)^2 - b^2 p^2}\right),
\eeq
and one of hamilton's equations gives the velocity as
\beq
u^\mu = - {\partial {\cal H}^* \over \partial p_\mu} 
= {e \over m} \left( p^\mu \mp {(b \cdot p)b^\mu - b^2 p^\mu \over 
\sqrt{(b \cdot p)^2 - b^2 p^2}}\right),
\eeq
which can be inverted to give
\beq
p^\mu = {m u^\mu \over e} \pm {(u \cdot b) b^\mu - b^2 u^\mu \over \sqrt{(b \cdot u)^2 - b^2 u^2}},
\eeq
providing a well-defined Legendre transformation between the lagrangian and extended 
hamiltonian functions, at least away from singular points.  
Behavior near singular points is discussed further in $\cite{colplb,colmcd}$.

\section{Quantization}
A typical field expansion contains the observer-covariant phase space factor
\beq
\int {d^3 \vec p \over 2 p^0(\vec p)} = \int d^4 p~ \delta (p^2 - m^2) \Theta(p^0),
\eeq
which is problematic in frames where space-like states satisfy $p^0 \le 0$ and the naive 
commutators
\beq
[a(p), a^\dagger(p)] = (2 \pi)^3 2 p^0(\vec p) \delta (\vec p - \vec p^\prime)
\eeq
vanish or go negative, leading to interpretational difficulties.
To circumvent this problem, the correct relativistic hamiltonian for the theory can be used
in the delta function to put the particles properly on-shell according to their classical mechanical
limit.
\beq
{\cal H}^*_\pm 
= - {e \over 2 m} \left( p^2 - m^2 - b^2 \pm 2 \sqrt{(b \cdot p)^2 - b^2 p^2}\right),
\eeq
so that the phase-space integral becomes
\beq
\int d^4 p ~ {-e \over 2 m}\delta ({\cal H}_\pm^*(p))  = \int {d^3 \vec p \over \La_\pm^{0 \prime}(p)}
\eeq
with
\beq
\La_\pm^{0 \prime}(p) = - \left( {2m \over e} \right){\partial{\cal H}^*_\pm \over \partial p^0} 
= 2 \left( p^0 
\pm {(b \cdot p) b^0 - b^2 p^0 \over \sqrt{(b \cdot p)^2 - b^2 p^2}} \right),
\eeq
which enforces the correct dynamics on surface ${\cal H}_\pm^* = 0$.
Since $u^2 =u_0^2 - \vec u^2 =1$ (in proper time parameterization), 
the factor $u^0 = - \partial {\cal H}^*_\pm / \partial p^0  > 0$ in all frames
so denominator in the phase-space factor is positive definite.
Using this factor in the field expansions (with the appropriate photon extended hamiltonian
\cite{covphotquant}) yields
\beq
A^\mu(x) = \int \sum_\pm {d^3 \vec p \over \La_\pm^{0 \prime}(p)} 
\ep^\mu a^\dagger (p) e^{-i p \cdot x} + c.c.,
\eeq
and the corresponding phase space factors are well-defined in all frames, not just concordant 
\cite{ralph}
ones in which the Lorentz-violating parameters are small.
Defining the canonical momenta $\pi^\mu = \partial {\cal L} / \partial \dot A_\mu$ gives
\beq
\pi^\mu(x) = F^{\mu 0}(x) + \epsilon^{0\mu\alpha\beta}(k_{AF})_\alpha A_\beta(x) - \eta^{\mu 0}\frac{1}{\xi}\partial_\nu A^\nu(x).
\eeq
Imposing the canonical quantization rules
\beq
\left[A_\mu(t,\vec{x}),\pi^\nu(t,\vec{y})\right] = i\delta^{\nu}_\mu \delta^3(\vec{x}-\vec{y})\ , \quad
\left[A_\mu(t,\vec{x}),A_\nu(t,\vec{y})\right] = 0,
\eeq
yields the momentum-space state algebra
\beq
[a(\vec p), a^\dagger (\vec p^\prime)] = (2 \pi)^3 \La_\pm^{0 \prime}(p) \delta (\vec p - \vec p^\prime),
\eeq
which has a positive-definite $\La_\pm^{0 \prime}(p) > 0$ in all observer frames.

\section{Summary}
Lorentz-violating field theories that lead to space-like states are not automatically ruled out
at the quantum level, even though there may exist
frames in which the energy runs negative.
By using an appropriate modified hamiltonian formalism, it is possible (at least in certain specific cases) to define positive-definite phase space factors that can be used to expand the fields
and impose a consistent separation between particle and anti-particle states.
The corresponding phase-space factors are important in consistent calculations like Cherenkov
processes \cite{potting1,potting2}.

\section{Acknowledgments}
I would like to thank the New College faculty development summer funds program for 
funding.

\end{document}